\documentclass[review,authoryear,12pt]{elsarticle}

 \usepackage{graphicx}
\usepackage{amssymb}
\journal{Icarus}

\begin{document}

\begin{frontmatter}

\title{Topographic Constraints on the Origin of the Equatorial Ridge on Iapetus}

\author[label1]{Erika J. Lopez Garcia}
\author[label1]{Edgard G. Rivera-Valentin}
\author[label2]{Paul M. Schenk}
\author[label1]{Noah P. Hammond}
\author[label1]{Amy C. Barr}

 \address[label1]{Department of Geological Sciences, Brown University, Providence, RI 02912}
 \address[label2]{Lunar and Planetary Institute, Houston, TX 77058}

\begin{abstract}
Saturn's moon Iapetus has an equatorial ridge system, which may be as high as 20 km, that may have formed by endogenic forces, such as tectonic and convective forces, or exogenic processes such as debris infall. We use high-resolution topographic data to conduct a topographic analysis of the ridge, which suggests a predominantly triangular morphology, with some ridge face slopes reaching 40$^{\circ}$,  allowing for an exogenic formation mechanism.
\end{abstract}

\begin{keyword}
Iapetus \sep  Satellites, surfaces \sep Geological processes

\end{keyword}

\end{frontmatter}

\section{Introduction}

Low resolution \emph{Voyager 2} images show that Saturn's outermost regular satellite, Iapetus, has unusually high topography \citep{Denk2000Iapetus}. Later, high-resolution \emph{Cassini} images show several of the peaks identified from \emph{Voyager 2}, and  illustrate that those peaks are part of an equatorial ridge system extending over 110$^{\circ}$ in longitude \citep{Porco:2005aa}. Previous studies report that parts of the ridge reach a height of 20 km (as determined from limb measurements) and could be up to 70 km wide \citep{Porco:2005aa, Giese:2008aa}.  Parts of the ridge are discontinuous and some sections have up to three nearly parallel ridge sections \citep{Porco:2005aa, Ip:2006aa, Giese:2008aa}. No significant feature pre-dating the ridge system has been discovered \citep{Porco:2005aa, Neukum:2005aa, Ip:2006aa, CastilloIapetus2007, Giese:2008aa}, attesting to its old age. 

The location of the ridge at Iapetus's equator may suggest an endogenic origin. Some suggest tectonism and volcanism as possible endogenic formation mechanisms. Despinning can create complex tectonic patterns on solid planetary bodies \citep{pechmann1979global, Porco:2005aa}; however, the tectonic features predicted by this mechanism have not been observed \citep{Porco:2005aa, CastilloIapetus2007, Singer:2011aa}. \citet{Giese:2008aa} suggest that the ridge may be tectonic because some slopes measured at the sides of the ridge are less than the angle of repose for ice, assuming the ridge material had been deposited. The possible tectonic activity could be related to the upwarping of the surface on either side of the ridge \citep{Giese:2008aa}. Others argue the appearance of the ridge is consistent with its formation by material rising from below. \citet{Denk:2005aa} suggest the ridge is a result of extrusive volcanic activity.  

Exogenic hypotheses argue that debris infall may have formed the ridge. \citet{Ip:2006aa} proposes that the ridge represents the remains of an early ring system that fell onto the equator. The resultant debris should have a slope close to the angle of repose, which may be between the values for two possible models of ridge material: rounded grains, for which the angle of repose is about $\sim$25$^{\circ}$ \citep{Kleinhans:2011aa}, or for snow mixed with particles of hail, for which the angle of repose $\sim 45^{\circ}$ \citep{Abe2004}. The cross-sectional morphology should show a single prominent peak and be uniform as a function of longitude \citep{Ip:2006aa}; however, later impact modification would change the morphology of the ridge. \citet{Levison:2011aa} suggest a scenario in which an ancient impact formed a sub-satellite and an impact-generated disk around Iapetus. The remains of the disk fell onto the surface to form the ridge, and the sub-satellite later impacts Iapetus to form one of its large basins during its tidal evolution \citep{Levison:2011aa}. \citet{Dombard:2012Iapetus} also suggest the ridge formed from a sub-satellite. They propose that tidal forces decay and tear the sub-satellite apart, and its debris falls onto the equator. The discovery of a possible ridge system on Rhea \citep{Schenk:2011aa}, may suggest that these types of events are common \citep{Dombard:2012Iapetus}.

Past characterization of the ridge (e.g., \citealt{Giese:2008aa}) has been limited by the number and quality of digital elevation models (DEMs) available. Characterizing the ridge slopes and morphology is key to deciphering the origin of the ridge. Here, we use DEMs generated from \emph{Cassini} imagery \citep{SchenkDPS2010} to observe how the ridge slopes and cross-sectional morphology varies along its length.

\section{Methods}
\subsection{Profiles}
We use eight \emph{Cassini} Imaging Science Subsystem (ISS) images with corresponding DEMs, which are constructed by a combination of stereo and photoclinometry, of the ridge system within Iapetus' dark terrain to obtain topographic profiles. We focus our morphological study on the dark terrain, where the topography of the ridge is most readily calculated \citep{SchenkDPS2010, White2013}. \emph{Cassini} ISS coverage of the equatorial ridge, as well as other parts of Iapetus, was used to produce these topographic maps \citep{SchenkDPS2010}. The images range in size and resolution, allowing us to study the ridge on both a local and regional scale. 

The wide range of spatial coverage allows observations of the ridge as a semi-continuous system across many kilometers and to closely observe its morphology. We obtained a total of 506 topographic profiles across all eight DEMs.

Figure 1 illustrates the footprints of the study images and the locations of the topographic profiles. Depending on the size of the region covered by each DEM, profiles were taken between $-10^{\circ}$ to $10^{\circ}$ latitude, which is about 250 km. There was consistent spacing between the profiles, with a maximum of $0.1^{\circ}$ latitude. Profiles were taken perpendicular to the ridge and include both the ridge and adjacent plains.  Three points on the ridge, including the peak and an endpoint, are used to divide the ridge topography into four segments: northern and southern ridge faces and flanks.  For example, in the trapezoidal ridge example in Fig. 2b, the transition from northern flank to face occurs at about $2^{\circ}$ latitude and the transition from southern face to flank is close to $-4^{\circ}$ latitude. Least squares fitting was used to determine the best fit line to the topography and corresponding slope angle of each segment with its corresponding standard error of the mean (SEM) to a 95\% confidence. Error bars reported here are upper bound values and encompass propagating any errors from lighting or viewing geometry effects. We report positive slopes as north-facing slopes and negative slopes as south-facing slopes.
 
\subsection{Morphology classification}
Profile segments that were long enough to cover both the ridge and the adjacent plains are classified based on its morphology.  We identify six morphology types defined by the shape of the peak, number of peaks, and the number of different types of peaks found in a single profile: triangular, trapezoidal, crowned, twinned, dissimilar and saddle. An example profile for each morphology is shown in Figure 2.

A triangular peak, as observed by previous work \citep{CastilloIapetus2007, Giese:2008aa}, is characterized by its distinct sharp peak with smooth sides on either face of the ridge. The trapezoidal morphological type described by \citet{Giese:2008aa} has a pronounced tabular top, similar to the famous Table Mountain in Cape Town, South Africa \citep{luckoff1951table}. A crowned peak, which has been previously described as a group of three parallel ridges \citep{Porco:2005aa, Ip:2006aa, Giese:2008aa}, is characterized by three triangular peaks, with the middle peak being taller than the two flanking peaks. A twinned morphology has two peaks side by side, where each peak has a similar shape. In some cases, we find two different morphology types in a single profile -- often, one triangular peak and one smaller trapezoidal peak -- which we name ``saddle'' peaks. Dissimilar peaks are peaks that have at least three peaks of different sizes and shapes. We also find some profiles where the ridge wasn't obviously visible or had no distinct morphology.

\section{Results and Discussion}
We find the triangular peak is the most common morphology, observed in 33\% of all of the profiles with a distinguishable shape. For this morphology, the average slopes for the northern and southern faces are similar to previous results, $\sim 15^{\circ}$ \citep{Giese:2008aa}, ranging from $\sim 3^{\circ}$ to $\sim 39^{\circ}$ (see Table 1). The trapezoidal peak morphology is the second most frequent, seen in 21\% of profiles. Saddle peaks occur in 17\% of the profiles and also have face slopes similar to triangular peaks. Twinned peaks occur in 14\% of the profiles and crowned peaks are observed in 8\% of the profiles. Lastly, the dissimilar ridge morphology is the least common morphology (7\% of profiles). 

Because we find the triangular shape to have the steepest slopes, and the sharpest peaks, our results suggest the triangular shape could be the least impact-modified ridge morphology on Iapetus. Trapezoidal ridge morphologies could be the result of landslides (e.g., \citealt{Singer:2012aa}), or impact modification. We also suggest that twinned, crowned, and dissimilar peaks formed by further impact modification or landslides \citep{Singer:2012aa}. The frequency of trapezoidal and saddle ridge morphologies are similar (21\% and 17\%), possibly indicating that trapezoidal and saddle peaks represent degraded triangular peaks, which could be modified by landslides \citep{Singer:2012aa}. The other less frequent categories have extremely shallow slopes, $<13^{\circ}$, suggesting significant erosion.  Generally, there is not a typical transition between morphologies. 

We find the flanks adjacent to the ridge have a significant slope. Specifically, the southern flank slopes for triangular, trapezoidal, crowned, and saddle peaks are high, $\sim5^{\circ}$. This may indicate that the ridge system extends further out than previously thought \citep{Porco:2005aa, Giese:2008aa}, may be built upon a pre-existing uplift or bulge, and/or may be more massive than originally estimated. There seems to be no sharp boundary between the faces of the ridge and the plains.

\section{Conclusions}
Using high-resolution imagery data, we conduct a topographic survey of the equatorial ridge of Saturn's moon Iapetus. We obtained 506 topographic profiles across eight \emph{Cassini} ISS images with corresponding DEMs to calculate the slope of the northern and southern faces and flanks of the ridge.

In addition to the three morphological types already classified - triangular, trapezoidal, and crowned peaks - \citep{Porco:2005aa, CastilloIapetus2007, Giese:2008aa}, we find three additional ridge shapes: a twinned morphology, a dissimilar morphology, and a saddle morphology. Previous work had ruled out exogenic formation processes due to shallow ridge slopes and varied morphology \citep{Giese:2008aa}; however, we find that the most common morphology of the ridge is a triangular peak with face slopes reaching $\sim$40$^{\circ}$. Slopes on triangular ridge faces are close to the angle of repose, which can vary between $\sim 45^{\circ}$ for snow with small particles of hail \citep{Abe2004} or $\sim$25$^{\circ}$ for rounded grains   \citep{Kleinhans:2011aa}, for $\sim$10\% of the topographic profiles. This may point to an exogenic origin for the ridge, such as formation by debris infall. 

Iapetus is among the most unusual planetary bodies in the Solar System. Although the formation of its equatorial ridge remains mysterious, the most common triangular ridge morphology and the evidence of slope angles close to the angle of repose make the case for an exogenic origin more plausible. 

\section*{Acknowledgements}
Lopez Garcia acknowledges support from the Mellon Mays Undergraduate Fellowship. This work was supported by NASA OPR NNX12AL22G and PGGNNX08AKA3G.  We thank A. Nahm and K. Singer for helpful reviews.

\section*{References}



\clearpage

\vspace{0.25in}
\begin{table}[htbp]

  \centering
    \begin{tabular}{l l l l l l}
    \hline
          & \textbf{North Flank} & \textbf{North Face} & \textbf{South Face} & \textbf{South Flank} \\
    \hline
    \textbf{Triangular} &       &       &       &  \\
    \textit{average} & $3.0^{\circ} \pm 1.1^{\circ}$ & $15.2^{\circ} \pm1.4^{\circ}$ & $-16.5^{\circ} \pm 1.7^{\circ}$ & $-4.0^{\circ} \pm 1.5^{\circ}$ \\
    \textit{range} & $-9.9^{\circ}$ to $13.8^{\circ}$ & $3.8^{\circ}$ to $29.4^{\circ}$ & $-38.8^{\circ}$ to $-2.7^{\circ}$ & $-19.4^{\circ}$ to $6.9^{\circ}$ \\
    \textbf{Trapezoidal} &       &       &       &  \\
    \textit{average} & $3.8^{\circ} \pm 1.0^{\circ}$ & $15.9^{\circ} \pm 1.5^{\circ}$ & $-16.8^{\circ} \pm 1.8^{\circ}$ & $-4.0^{\circ} \pm 1.4^{\circ}$ \\
       \textit{range} & $-9.7^{\circ}$ to $11.3^{\circ}$ & $3.3^{\circ}$ to $29.2^{\circ}$ & $-39.9^{\circ}$ to $-5.7^{\circ}$ & $-16.2^{\circ}$ to $5.7^{\circ}$ \\
      \textbf{Crowned} &       &       &       &  \\
    \textit{average} & $3.8^{\circ} \pm 1.1^{\circ}$ & $13.7^{\circ} \pm 3.3^{\circ}$ & $-13.1^{\circ} \pm 3.6^{\circ}$ & $6.3^{\circ} \pm 2.2^{\circ}$ \\
      \textit{range} & $0.7^{\circ}$ to $10.7^{\circ}$ & $-9.0^{\circ}$ to $26.5^{\circ}$ & $-36.1^{\circ}$ to $7.1^{\circ}$ & $-13.9^{\circ}$ to $4.3^{\circ}$ \\
       \textbf{Twinned} &       &       &       &  \\
    \textit{average} & $4.5^{\circ} \pm 1.4^{\circ}$ & $8.9^{\circ} \pm 3.1^{\circ}$ & $-8.0^{\circ} \pm 3.2^{\circ}$ & $-2.7^{\circ} \pm 1.3^{\circ}$ \\
    \textit{range} & $-5.5^{\circ}$ to $14.5^{\circ}$   & $-10.0^{\circ}$ to $23.7^{\circ}$ & $-29.7^{\circ}$ to $10^{\circ}$ & $-13.3^{\circ}$ to $7.2^{\circ}$ \\
       \textbf{Dissimilar} &       &       &       &  \\
        \textit{average} & $3.1^{\circ} \pm 2.7^{\circ}$ & $11.9^{\circ} \pm 4.8^{\circ}$ & $-10.1^{\circ} \pm 3.9^{\circ}$ & $-2.8^{\circ} \pm 1.7^{\circ}$ \\
    \textit{range} & $-12.2^{\circ}$  to $12.9^{\circ}$ & $-16.0^{\circ}$ to $32.3^{\circ}$ & $-22.3^{\circ}$ to $8.2^{\circ}$ & $-9.2^{\circ}$ to $5.7^{\circ}$ \\
    \textbf{Saddle} &       &       &       &  \\
    \textit{average} & $4.3^{\circ} \pm 1.2^{\circ}$ & $14.8^{\circ} \pm 1.5^{\circ}$ & $14.6^{\circ} \pm 1.9^{\circ}$ & $-5.6^{\circ} \pm 1.7^{\circ}$\\
    \textit{range} & $-5.3 ^{\circ}$ to $14.6^{\circ} $ & $1.3^{\circ} $ to $22.3^{\circ} $ & $-33.9^{\circ} $ to $-3.2^{\circ} $ & $-16.3^{\circ} $ to $4.4^{\circ} $ \\

     \hline
    \end{tabular}%
    \caption{\small{Average and range of slope angles for the six ridge morphology types. }}
  \label{tab:slope.results}%
   
\end{table}%


\clearpage

\begin{figure}\label{fig:global.map}
 \centerline{\includegraphics[scale=0.3]{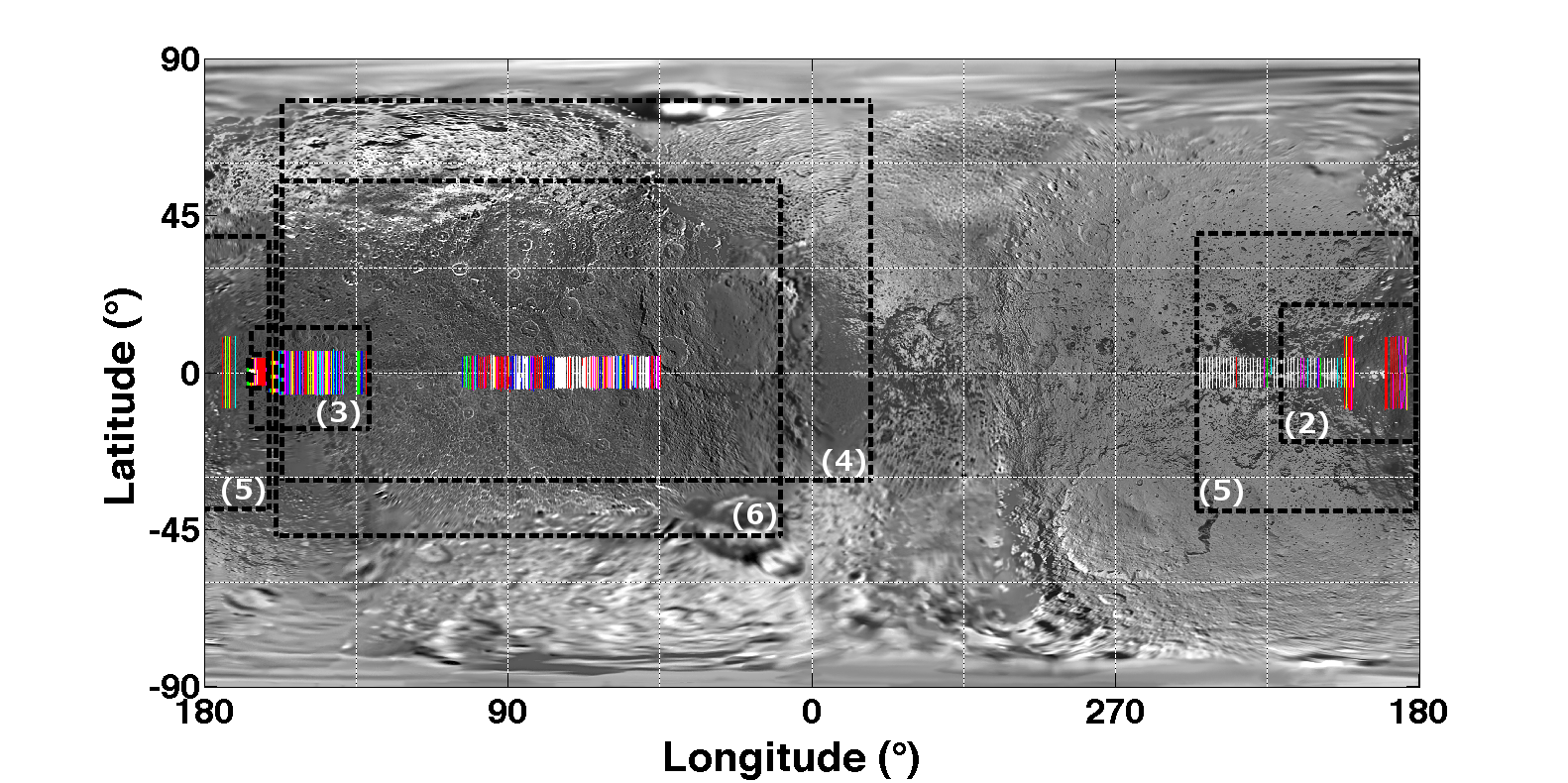}}
      \caption{\small {A global map of Iapetus (image PIA11116 resolution of 0.8 km/pxl) with the location of each image (except images 1, 7 and 8) and the types of ridge morphology observed. Red is triangular peak, pink is trapezoidal peak, green is crowned peak, blue is twinned peak, cyan is dissimilar peak, yellow is saddled peak, and white is no discernible peak. These are the image names that are on the global map: (2) W1568130503 (resolution 0.41 km/pxl); (3) W1568125439 (resolution 0.32 km/pxl); (4) N1483151512 (resolution 0.90 km/pxl); (5) W1568133373 (resolution 0.79 km/pxl); and (6) N1482859934 (resolution 6.7 km/pxl). Images (1) W1568127165 (resolution 0.09 km/pxl, latitude and longitude $-4.48083^{\circ}$, $160.992^{\circ}$, $5.21678^{\circ}$, $167.332^{\circ}$),  (7) N1568127472 (resolution 0.009 km/pxl, latitude and longitude $-0.502969^{\circ}$, $165.907^{\circ}$, $0.65417^{\circ}$, $166.711^{\circ}$), and (8) N1568127660 (resolution 0.809 km/pxl, latitude and longitude $-0.36639^{\circ}$, $165.015^{\circ}$, $0.65908^{\circ}$, $165.939^{\circ}$) are too small to be depicted on this map.}}  
\end{figure} 

\clearpage

\begin{figure}\label{fig:ridge.morph.ex}
 \centerline{\includegraphics[scale=0.35]{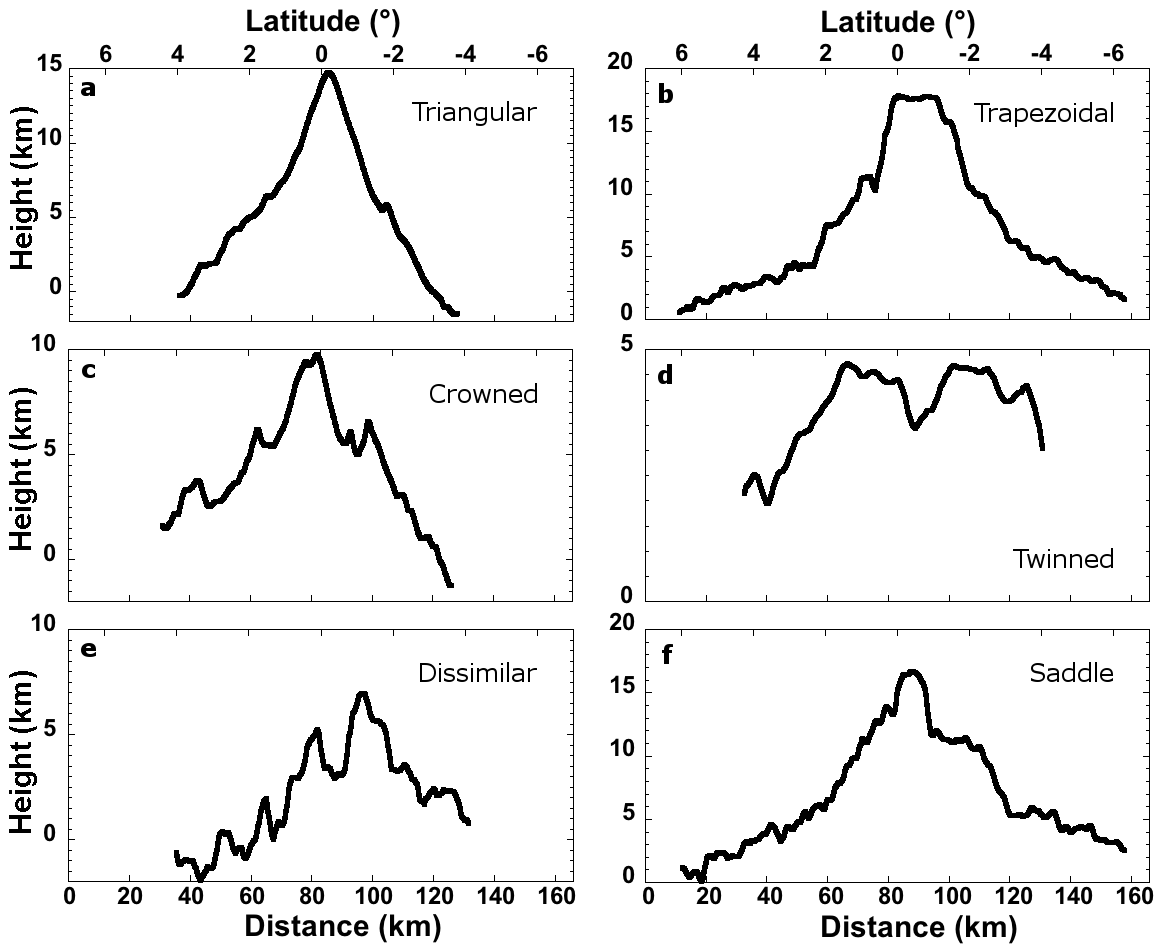}}
      \caption{\small Representative examples of the six ridge morphological types observed in the topographic profiles. Vertical exaggeration $\sim 10$ times.} 
\end{figure}

\end{document}